\documentstyle[twoside,fleqn,espcrc2,epsf]{article}

\newcommand{\BE}{\begin{equation}}
\newcommand{\EE}{\end{equation}}
\newcommand{\BEA}{\begin{eqnarray}}
\newcommand{\EEA}{\end{eqnarray}}
\newcommand{\BA}{\begin{array}}
\newcommand{\EA}{\end{array}}

\title{Computing the lowest eigenvalues of the Fermion matrix
       by subspace iterations}

\author{B. Bunk
      \address{Institut f{\"u}r Physik, Humboldt--Universit{\"a}t zu Berlin,
            Invalidenstr.110, D--10115 Berlin, Germany}
      \thanks{email: bunk@linde.physik.hu-berlin.de}
      }

\begin{document}

\begin{abstract}
Subspace iterations are used to minimise a generalised Ritz functional of
a large, sparse Hermitean matrix.
In this way, the lowest $m$ eigenvalues are determined.
Tests with $1 \leq m \leq 32$ demonstrate that
the computational cost (no. of matrix multiplies)
does not increase substantially with $m$. This implies that,
as compared to the case of a $m=1$,
the additional eigenvalues are obtained for free.
\end{abstract}

\maketitle

\section{Introduction}

For a large, sparse Hermitean matrix $A \in C^{N \times N}$,
minimisation of the Ritz functional
\BE
      q(x) = \frac{x^\dagger Ax}{x^\dagger x} \quad , \quad x \in C^N .
\EE
provides the smallest eigenvalue, together with the corresponding
eigenvector. The minimum of a functional can be found iteratively by the
method of conjugate gradients (CG)\cite{NR}, its application to the special
case of the Ritz functional has been worked out by Geradin\cite{Gerad71}
and Fried\cite{Fried72}.

If a small number of lowest eigenvalues is required instead, $q(x)$ may
be minimised repeatedly, restricting $x$ to the space orthogonal to the
previous eigenvectors. An efficient implementation of this idea is
described in \cite{Kalk95}.

\section{Subspace iterations}

An alternative approach to compute the $m$ smallest eigenvalues (with eigenvectors)
simultaneously considers the corresponding subspace as a whole.
An $m$--dimensional subspace of $C^N$ is spanned by $m$ non--degenerate (column)
vectors, which are combined into a rectangular matrix $x \in C^{N \times m}$.
From now on, $N$ is considered `large', $10^5$ say, but $m$ is `small',
e.g. $O(10)$.

The projector onto the subspace is
\BE
      P(x) = x \, (x^\dagger x)^{-1} x^\dagger .
\EE
Note that the inversion of $x^\dagger x \in C^{m \times m}$ is a small problem
and can be solved by standard techniques.
Minimisation is now applied to the generalised Ritz functional\cite{Meyer87}
\BE
      q(x) = Tr\{P(x)A\} .
\EE
At its minimum, $P(x)$ projects onto the space of the $m$ lowest eigenvalues.

An iterative CG minimisation of $q(x)$ starts with the computation of the
gradient at $x$:
\BE
      g = (1 - P(x))\, A x \, (x^\dagger x)^{-1} \quad  \in  C^{N \times m} .
\EE
By construction, $x^\dagger g = 0$. A steepest descent method would search a new
minimum along $g$, but CG algorithms use an ansatz with a more general shift
`direction' $h \in C^{N \times m}$ (to be discussed later):
\BE
      x' = x + h \alpha       \label{eq-x'}
\EE
with a (small) coefficient matrix $\alpha \in C^{m \times m}$.
It is helpful to assume
\BE
      x^\dagger h = 0   ,        \label{eq-xh}
\EE
i.e. all shift vectors (columns of $h$) are orthogonal to the
current subspace (spanned by the columns of $x$). The matrix
$\alpha$ is to be determined to minimise $q(x + h\alpha)$. To bring this
to a manageable form, consider the projection of $A$ into the space spanned by
the columns of $x$ and $h$. This defines a $(2m)$--dimensional eigenvalue problem
(in a non--orthogonal basis), which is to be solved for the $m$ lowest eigenvalues.
The Jacobi method\cite{NR} can deal with such a `small' problem.
In a matrix notation using $m \times m$ blocks, the result can be expressed as
\BEA
      \left( \BA{cc}    x^\dagger A x  & x^\dagger A h      \\
                        h^\dagger A x  & h^\dagger A h      \EA \right)
      \left( \BA{c}     \beta_1     \\
                        \beta_2     \EA \right)
                              \quad\quad\quad\quad    \nonumber \\
      =
      \left( \BA{cc}    x^\dagger  x   & 0               \\
                        0           & h^\dagger  h       \EA \right)
      \left( \BA{c}     \beta_1     \\
                        \beta_2     \EA \right) \Lambda .
\EEA
The eigenvalues are in $\Lambda = \mbox{diag}(\lambda_1 \ldots \lambda_m)$
and the eigenvectors are the columns of
\BE
      \left( \BA{c}     \beta_1     \\
                        \beta_2     \EA \right) \in C^{2m \times m} .
\EE

The new minimising subspace is spanned by $x \beta_1 + h \beta_2$. Comparision
with eq.(\ref{eq-x'}) reveals that the minimum of $q(x + h\alpha)$ is found at
\BE
      \alpha = \beta_2 \beta_1^{-1} .
\EE

\section{Update of the search space}

Following the general CG scheme, one starts with $h = g$, but subsequently
$h$ is updated to a superposition of the actual gradient and the previous
search direction. An appropriate ansatz is
\BE
      h' = g' + (1 - P(x')) \, h \gamma
\EE
($g'$ is the gradient evaluated at $x'$). The projection in front of $h$
insures the orthogonality eq.(\ref{eq-xh}) for $h'$.
$\gamma \in C^{m \times m}$ is another (small) matrix.
Its choice is somewhat arbitrary: in the linear case (minimisation of a quadratic
form), it is determined by the requirement that the search vectors should be
conjugate, but this argument does not carry over to the nonlinear situation.
Experience has shown that the following form 
\BE
      \gamma = (g^\dagger g)^{-1} \left\{ g'^\dagger g' - g^\dagger g' \right\}
\EE
is reasonably efficient and robust. It is a matrix version of the
Polak--Ribiere prescription\cite{NR}. Note that matrices like
$g^\dagger g \in C^{m \times m}$ are small.

\section{Convergence and stopping}

\begin{figure}[t]
\caption{Relative shifts of the four lowest eigenvalues}
      \label{fig-shifts}
\noindent
\epsfxsize=75mm
\leavevmode
\epsffile{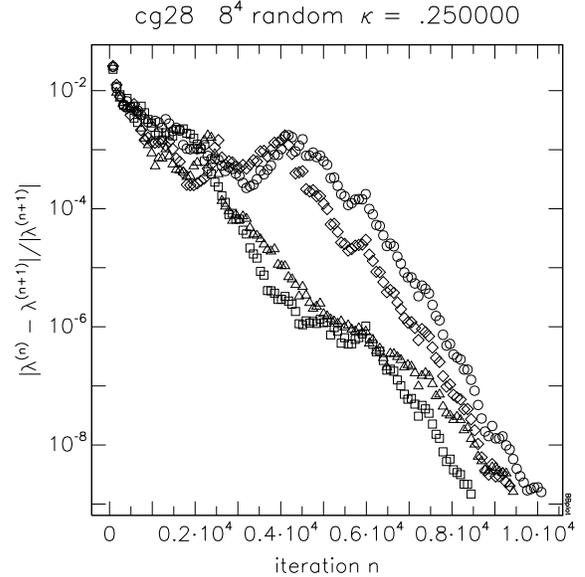}
\end{figure}

By construction, $q(x)$ will decrease monotonically with the iteration number
$n$, but the rate of convergence and the behavior of individual eigenvalues
is fairly irregular and errors have to be estimated with care. As an
example, Fig. \ref{fig-shifts} shows the relative shifts of the four lowest
eigenvalues of the fermion matrix $M_f^\dagger M_f$ for Wilson fermions on an
$8^4$ random $SU(2)$ gauge configuration.

The following procedure is proposed: assuming geometric convergence
\BE
      q(x^{(n)}) \approx q(x^{(\infty)}) + a f^n ,
\EE
$f$ is computed from three {\em widely separated} $n$'s.
Then the relation
\BE
      q(x^{(n)}) - q(x^{(\infty)}) \approx
                  \frac{1}{1-f} [q(x^{(n)}) - q(x^{(n+1)})]
\EE
shows how `local' shifts are to be rescaled to provide error estimates:
an eigenvalue $\lambda_i$ is considered {\em converged} within relative error
$\delta$ if
\BE
      \frac{1}{1-f} |\lambda_i^{(n)} - \lambda_i^{(n+1)}|
                  < \delta \, |\lambda_i^{(n+1)}| .
\EE
The corresponding eigenvector is frozen and only used in further iterations to
keep the remaining subspace orthogonal to it.

\begin{figure}[t]
\caption{Relative errors of the four lowest eigenvalues}
      \label{fig-errors}
\noindent
\epsfxsize=75mm
\leavevmode
\epsffile{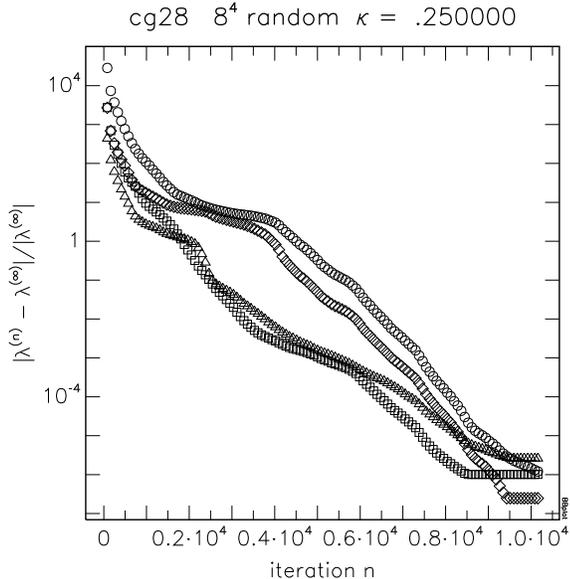}
\end{figure}

For the example given above, Fig. \ref{fig-errors} demonstrates that this
procedure leads to correct stopping as the eigenvalues approach the precise
values within a prescribed relative error of $\delta = 10^{-6}$.

\section{Computational cost}

Subspace iterations require to store the large arrays $x$, $h$, $Ax$, and
one more work matrix, this amounts to $4mN$ variables in total.

As with many sparse matrix algorithms,
the number of multiplications $Av$ (with a single column vector $v \in C^N$)
is considered a fair measure for the cost in cpu time. To estimate the
performance of the subspace iterations, tests with realistic matrices have
been performed: Wilson fermions in four dimensions with random $SU(2)$
gauge fields and (critical) hopping parameter $\kappa = 0.25$. In this case,
$A \sim M_f^\dagger M_f$.
For lattice sizes $4^4, 6^4, 8^4$, $m = 1,2,4,8,16,32$ eigenvalues and a relative
error bound $\delta = 10^{-6}$, the numbers of operations $Av$ needed for 
{\em all} eigenvalues to converge are shown in Table \ref{tab-ops}.

\begin{table}
\caption{Number of $Av$ operations to obtain $m$ lowest eigenvalues
      with precision $\delta = 10^{-6}$ on three random SU(2) lattices}
\label{tab-ops}
\begin{tabular}[t]{rccc}
\hline
m     & $4^4$     & $6^4$     & $8^4$     \\
\hline
1     & 2992      & 28500     & 29966     \\
2     & 2816      & 11911     & 40075     \\
4     & 4011      & 24414     & 37521     \\
8     & 4921      & 19322     & 50947     \\
16    & 5268      & 21494     & 64801     \\
32    & 6000      & 25089     & 66994     \\
\hline
\end{tabular}
\end{table}

The absolute number of iterations needed to achieve a given precision grows
with the matrix size (density of the eigenvalues), but the behavior is
difficult to understand in detail. Since diagonalisation
{\em within} the subspace is done non--iteratively, one might speculate that
the level spacings {\em around the $m$--th eigenvalue} are crucial.
This requires further studies.

In any event, the counts in Table \ref{tab-ops} show one encouraging feature:
the number of $Av$ operations does not grow substantially with $m$. 
The $Av$ part is due to one computation of $Ah$ per iteration and costs
$\sim Nm$. It will be dominated for larger $m$ by the
computation of small matrices like $x^\dagger x$ ($\sim N m^2$ per iteration),
but these scalar products are much cheaper and can be neglected for the moderate
values of $m$ considered.


\begin{thebibliography}{99}

\bibitem{NR}      W.~H.~Press, B.~P.~Flannery, S.~A.~Teukolsky, W.~T.~Vetterling,
                        Numerical Recipes, Cambridge University Press,
                        Cambridge, 1989.
\bibitem{Gerad71} M.~Geradin, J. Sound Vibration 19 (1971) 319.
\bibitem{Fried72} I.~Fried, J. Sound Vibration 20 (1972) 333.
\bibitem{Kalk95}  T.~Kalkreuter, H.~Simma, An Accelerated Conjugate Gradient
                        Algorithm to Compute Low--Lying Eigenvalues,
                        hep-lat/9507023, to appear in Comp. Phys. Comm.
\bibitem{Meyer87} A.~Meyer, Modern Algorithms for Large Sparse Eigenvalue
                        Problems, Akademie--Verlag, Berlin, 1987.
\end{thebibliography}
\end{document}